\pdfoutput=1

\documentclass[11pt]{article}

\usepackage[]{acl}
\usepackage{natbib}
\usepackage{times}
\usepackage{subcaption}
\usepackage{latexsym}
\usepackage{booktabs}
\usepackage{fancyvrb}
\usepackage{enumitem}
\usepackage{float}
\usepackage[T1]{fontenc}
\usepackage[utf8]{inputenc}

\usepackage{microtype}

\usepackage{inconsolata}

\usepackage{graphicx}
\usepackage{xcolor}  
\usepackage[utf8]{inputenc} 
\usepackage[T1]{fontenc}    
\usepackage{hyperref}       
\usepackage{url}            
\usepackage{amsfonts}       
\usepackage{nicefrac}       
\usepackage{microtype}      
\usepackage{graphicx}
\usepackage[most]{tcolorbox}
\usepackage{colortbl}
\tcbuselibrary{listings}
\usepackage{listings}
\newcommand{\sw}[1]{\textcolor{black}{#1}}

\usepackage[linesnumbered,ruled,vlined]{algorithm2e}

\definecolor{codegreen}{rgb}{0,0.6,0}
\definecolor{codegray}{rgb}{0.5,0.5,0.5}

\definecolor{backcolour}{RGB}{245,248,250}
\definecolor{emph}{RGB}{166,88,53}
\definecolor{nightblue}{RGB}{9,49,105}
\definecolor{keywords}{RGB}{207,33,46}
\definecolor{lightpurple}{RGB}{130,81,223}

\lstdefinestyle{mystyle}{
    backgroundcolor=\color{backcolour},   
    commentstyle=\color{codegreen},
    keywordstyle=\color{keywords},
    stringstyle=\color{nightblue},
    basicstyle=\fontsize{7}{8}\ttfamily,
    breakatwhitespace=true,         
    breaklines=true,                 
    captionpos=b,                    
    keepspaces=true,                 
    numberstyle=\tiny\color{codegray},
    numbersep=2pt,                  
    showspaces=false,                
    showstringspaces=false,
    showtabs=false,                  
    tabsize=2,
    emph={dsp,Example,sample,annotate,knn,crossval,generate,retrieve,retrieve\_ensemble,majority,fused_retrieval,Template, Transformation,rank,branch},
    emphstyle={\color{lightpurple}},
    linewidth=0.98\columnwidth,
    frame=tb,    
    xrightmargin=0pt,
    xleftmargin=0.23cm,
    numbers=left,
    aboveskip=0.4cm,
    belowskip=0.4cm,
}

\title{SEW: Self-Evolving Agentic Workflows for Automated Code Generation}

\author{
Siwei Liu$^{1}$ \quad
Jinyuan Fang$^{2}$ \quad
Han Zhou$^{3}$ \quad
Yingxu Wang$^{4}$ \quad
Zaiqiao Meng$^{2}$\Thanks{~Corresponding author.} \\
$^{1}$University of Aberdeen \quad
$^{2}$University of Glasgow \quad
$^{3}$University of Cambridge \quad
$^{4}$MBZUAI \\
\texttt{siwei.liu@abdn.ac.uk, hz416@cam.ac.uk, yingxv.wang@gmail.com,} \\
\texttt{\{jinyuan.fang,zaiqiao.meng\}@glasgow.ac.uk}
}

\def \OURS{SEW}

\begin{document}
\maketitle
\begin{abstract}
Large Language Models (LLMs) have demonstrated effectiveness in code generation tasks. To enable LLMs to address more complex coding challenges, existing research has focused on crafting multi-agent systems with agentic workflows, where complex coding tasks are decomposed into sub-tasks, assigned to specialized agents. 
Despite their effectiveness, current approaches heavily rely on hand-crafted agentic workflows, with both agent topologies and prompts manually designed, which limits their ability to automatically adapt to different types of coding problems. To address these limitations and enable automated workflow design,
we propose \sw{\textbf{S}elf-\textbf{E}volving \textbf{W}orkflow (\textbf{\OURS{}})},
a novel self-evolving framework that automatically generates and optimises multi-agent workflows. 
Extensive experiments on \sw{three coding benchmark datasets, including the challenging LiveCodeBench}, demonstrate that our \OURS{} can automatically design agentic workflows and optimise them through self-evolution, bringing \sw{up to 12\% improvement on LiveCodeBench compared to using the backbone LLM only}. 
Furthermore, by investigating different \sw{representation schemes} of workflow, we provide insights into the optimal way to encode workflow information with text.
\end{abstract}

\section{Introduction}


Large Language Models (LLMs) have demonstrated remarkable capabilities in automatic code generation, enabling developers to translate natural language descriptions into executable programs~\cite{hong2023metagpt,liu2024your}. However, as coding tasks grow in complexity, relying on a single LLM instance (single-agent) to handle all aspects of code generation becomes increasingly challenging. 
To address this, recent studies have explored multi-agent systems~\cite{huang2023agentcoder,islam2024mapcoder} where multiple LLM-powered agents collaborate to solve intricate problems through structured workflows~\cite{hong2023metagpt}. These multi-agent systems decompose complex programming tasks into sub-tasks, assigning them to specialized agents with tailored prompts, enhancing execution and output quality.

\begin{figure}[t]
    \centering
    \includegraphics[width=\columnwidth]{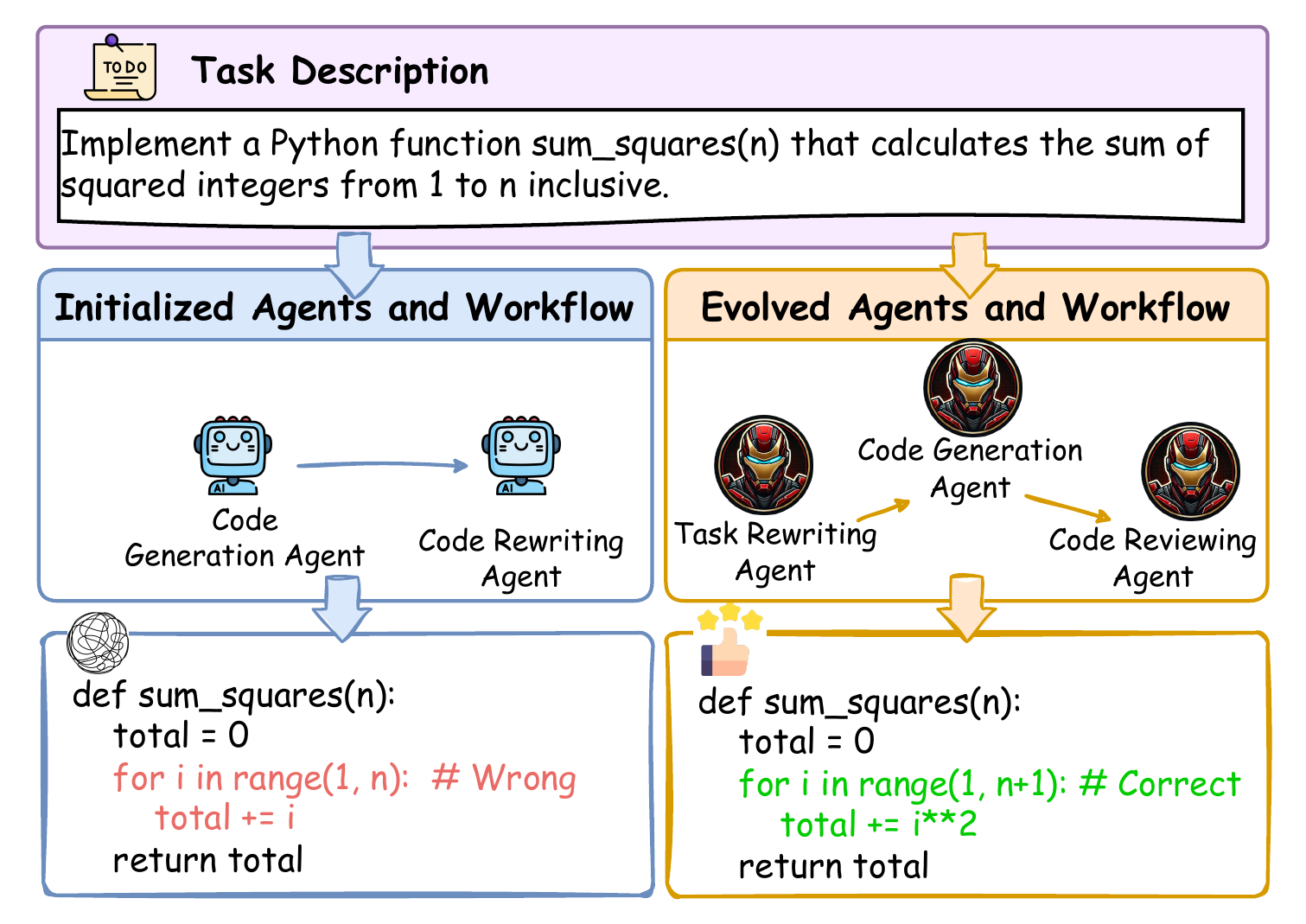}
    \vspace{-8mm}
    \caption{Illustration of agent and workflow evolution in code generation. \sw{The initialized setup (left) includes agents with naive prompts while the evolved setup (right) is equipped with enhanced prompts generated by \OURS{} and a more sophisticated workflow structure.}}
    \label{fig:trigger}
\end{figure}
Despite their effectiveness, current multi-agent 
systems rely heavily on manually designed workflows, where both the workflow topology and agents' prompts are \sw{manually crafted}, hindering their adaptability to \sw{more complex} \sw{coding task}. 
For instance, a workflow optimised for machine learning task~\cite{chi2024sela} differs significantly from one tailored for software development task~\cite{qian2023communicative}. Manually crafting workflows for each task is inefficient and does not leverage LLM’s full potential for autonomous adaptation.

To address these limitations, we propose \textbf{S}elf-\textbf{E}volving \textbf{W}orkflow (\textbf{\OURS{}}), a novel framework designed to automatically generate and optimise multi-agent workflow. 
\sw{In particular, \OURS{} achieves this by leveraging a novel evolutionary scheme to improve the workflow, i.e., the topology of workflows and the prompt of each agent.}
Figure \ref{fig:trigger} shows the agent and workflow evolution in code generation. In addition, to effectively represent agentic workflows in textual format, we explore and compare five different representation schemes, namely BPMN~\cite{white2004introduction}, CoRE~\cite{xu2024aios}, Python code~\cite{zhang2024pybench}, YAML~\cite{zhang2024effectiveness}, and pseudo-code~\cite{xiao2024flowbench}. We evaluate each scheme based on how well it can be interpreted and optimised by our \OURS{} framework, aiming to identify the optimal scheme for workflow representation and optimization. 
 
 Our contributions are: 
(1) We investigate different workflow representation schemes, such as BPMN, Python, CoRE, YAML, and pseudo-code, to determine the most effective format for LLM interpretation;
(2) \sw{Unlike prior work that builds agents by assembling predefined operators, our framework automatically constructs agentic workflows from scratch, conditioned solely on task descriptions.}
(3) We introduce a self-evolving workflow design approach, SEW, where LLMs \sw{jointly} improve workflow structures and agent prompts to optimise performance; 
(4) We conduct extensive experiments on three benchmark datasets, including MBPP, HumanEval+, and LiveCodeBench, demonstrating that \OURS{} can consistently improve workflow performance through self-evolution.

\section{Related Work}
\label{sec:related_work}

\subsection{Workflow Representations in Agents}
\label{subsec:workflow}

In multi-agent systems, workflows establish structured information flows and task execution pipelines, enabling agents to solve complex problems~\cite{hong2023metagpt, gao2024agentscope}. 
While natural language can describe workflows, its inherent ambiguity often leads to inconsistent interpretations, hindering precise task execution across agents~\cite{xu2024aios}.
To address this challenge, several studies have introduced specific representation schemes for SOPs. 
For example, \textit{Business Process Model and Notation} (BPMN)~\cite{white2004introduction} is a graphical modeling language designed to depict workflows by specifying the execution order of activities.
Similarly, \textit{Code Representation and Execution} (CoRE)~\cite{xu2024aios} provides a unified framework that integrates natural language programming, pseudo-code, and flow-based programming to improve workflow representation and execution.
Additionally, Python code~\cite{zhang2024pybench, xu2024theagentcompany}, YAML~\cite{qiao2023taskweaver, zhang2024effectiveness}, and pseudo-code~\cite{xiao2024flowbench, li2025structured} are also commonly employed to define and manage agentic workflows.

\begin{figure*}
    \centering
    \includegraphics[width=0.9\textwidth]{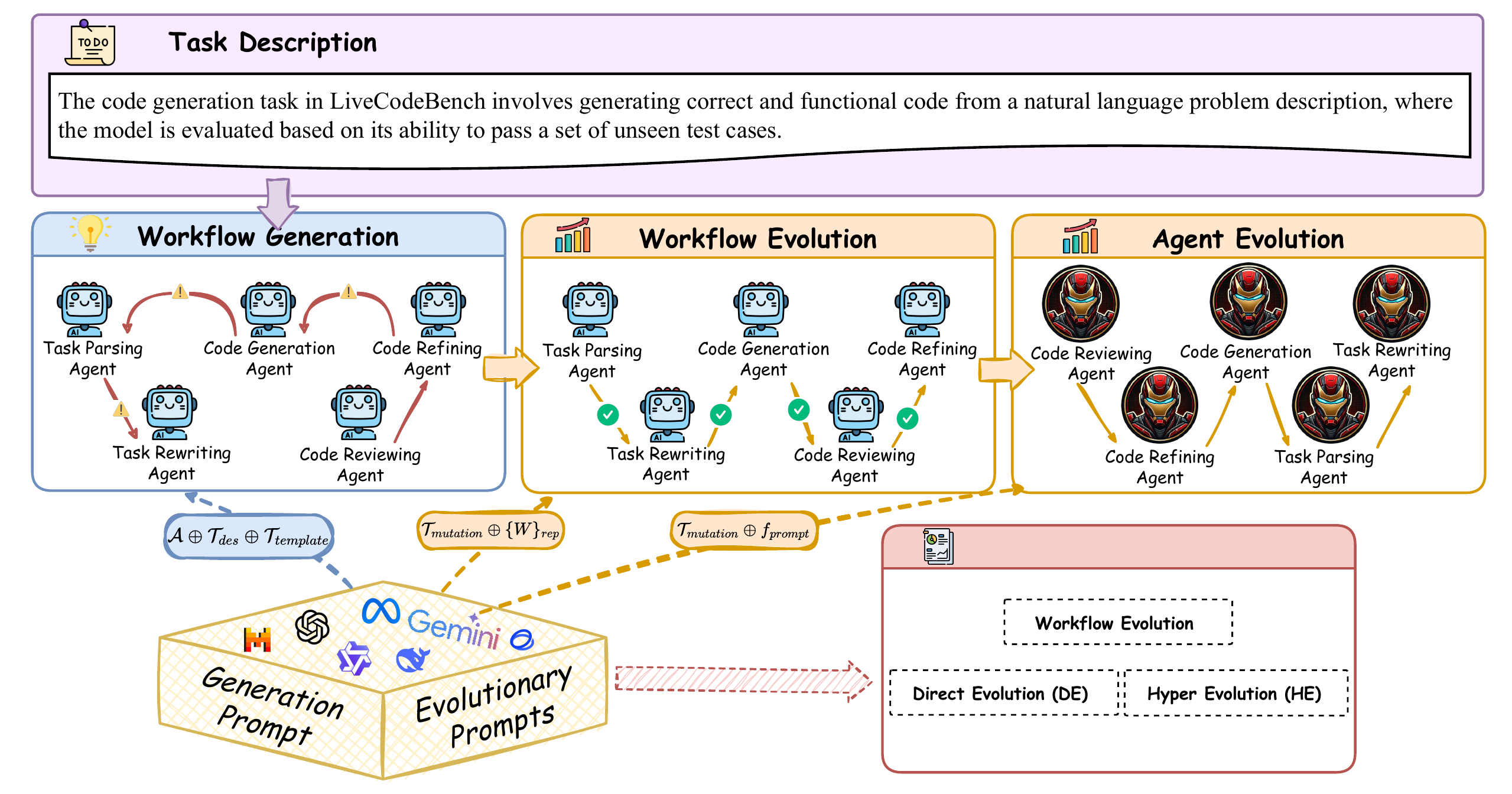}
    \vspace{-2mm}
    \caption{The overall framework of \OURS{}. The process begins with \sw{workflow generation}, followed by workflow \sw{evolution}. \sw{Then each agent within the evolved workflow will be equipped with enhanced prompts generated by the agent evolution module.}
    Such an agent evolution module is driven by the Direct Evolution (DE) operator and Hyper Evolution (HE) operator, leveraging LLMs, \sw{where we use a mutation prompt $\mathcal{T}_{mut}$ or a hyper-mutation prompt $\mathcal{T}_{hmut}$ to enhance the prompt of an agent}. }
    \label{fig:sew_framework}
\end{figure*}

\subsection{Self-Evolving Agents}
Existing agentic methods often yield suboptimal responses when prompts are poorly constructed.
To address this, prompt optimization techniques~\cite{zhoularge, promptbreeder, agarwal2024promptwizard, liu2024large} have moved beyond static, manually crafted in-context prompts. 
For instance, automatic prompt engineer~(APE)~\cite{zhoularge} enhances prompts by searching through a pool of candidates.
Similarly, Promptbreeder~\cite{promptbreeder} employs LLMs to mutate and evolve a population of task-specific prompts. \sw{MIPRO~\cite{opsahl2024optimizing} is an optimizer designed to enhance multi-stage language model programs by refining both instructions and few-shot examples for each module.}
In multi-agent systems, recent studies have explored the evolution of agentic workflows and topologies~\cite{zhang2024aflow,zhou2024symbolic,zhou2025multi,zhang2025evoflow}. 
For example, MASS~\cite{zhou2025multi} exploits the optimization of both prompt and workflow over a configurable topology space.
Similarly, AFlow~\cite{zhang2024aflow} employs a Monte Carlo Tree Search to enhance workflow efficiency, while EvoFlow~\cite{zhang2025evoflow} introduces a framework for the automated search of heterogeneous agentic workflows. \sw{EvoAgent~\cite{yuan2024evoagent} is designed to automatically extend expert agents into multi-agent systems using evolutionary algorithms.}
In contrast, our ~\OURS{} introduces a self-evolving mechanism that leverages diverse workflow representation schemes, jointly optimising prompts for both agents and their workflow.

\section{\OURS{}}\label{sec:sew}
\noindent \textbf{Task Definition.}
We focus on the task of code generation, a task that requires multi-agent collaboration~\cite{hong2023metagpt}, aiming to produce executable code based on a textual coding problem. To tackle this task, we deploy an LLM-based multi-agent system to generate code, where each agent processes a textual prompt and produces a corresponding textual output. \sw{We define the textual prompt of an LLM agent $a$ as $\mathcal{T}$ and a sequence of LLM agents, i.e., a workflow as $W$.}

\begin{figure*}
    \centering
    \includegraphics[width=\textwidth]{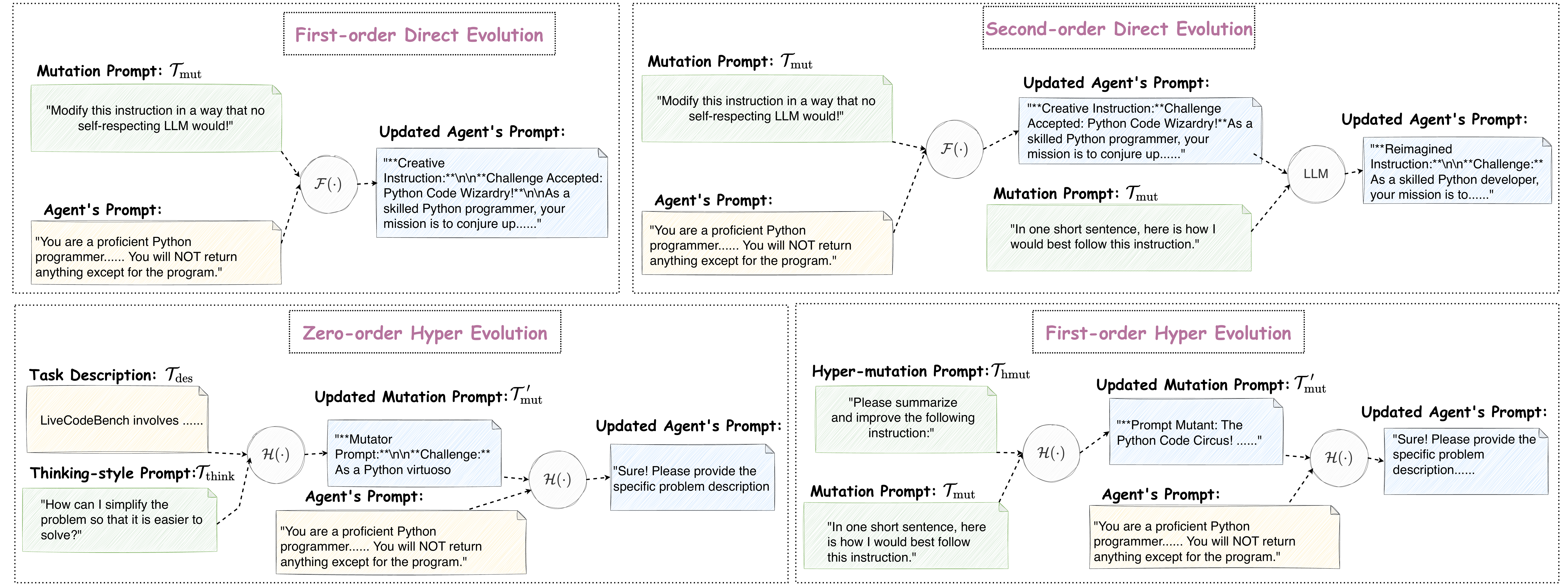}
    \caption{Illustration of the Direct Evolution and Hyper Evolution of \OURS{}. We use green, yellow and blue boxes to indicate the evolutionary prompt, default agent prompt and textual output of evolutionary operators.}
    \label{fig:evoprompt}
\end{figure*}

\noindent \textbf{Preliminary.}
Evolutionary prompts are central to \OURS{}. Rather than relying on training data, \OURS{} employs LLMs as mutation operators by concatenating the evolutionary prompts with the task prompt to generate a more effective task prompt. \sw{We define two evolutionary operators, namely the Direct Evolution (DE) operator $\mathcal{F}(\cdot)$ and the Hyper Evolution (HE) operator $\mathcal{H}(\cdot)$, where $\mathcal{F}(\cdot)$ and $\mathcal{H}(\cdot)$ take a workflow $W$ or an agent $a$ as the input and output an enhanced workflow $W'$ or an agent $a'$.} Specifically, \sw{$\mathcal{F}(\cdot)$ and $\mathcal{H}(\cdot)$ operators leverage (1) mutation prompts $\mathcal{T}_\text{mut}$, (2) hyper-mutation prompt, and (3) thinking-style prompts $\mathcal{T}_\text{think}$~\cite{promptbreeder}. Figure~\ref{fig:evoprompt} shows examples of these evolutionary prompts and how they are evolved by both DE and HE.}

\noindent \textbf{Overview of \OURS{}.} 
Our \OURS{} framework consists of three main modules: (a) Workflow \sw{Generation}, (b) Workflow-Evolution, and (c) Agent-Evolution. The overview of our \OURS{} framework is {illustrated} in Figure~\ref{fig:sew_framework}. \sw{As shown in Figure~\ref{fig:sew_framework}, our \OURS{} first generates an initial workflow based on the task description using one of the representation schemes introduced in Section~\ref{sec:rep}. Second, the workflow evolution module of \OURS{} will leverage our evolution method to reconstruct the initial workflow. Finally, inspired by PromptBreeder~\cite{promptbreeder}, our agent evolution module will apply either the agentic DE or agentic HE method to equip each agent with a more sophisticated prompt. The pseudo-code of \OURS{} is shown in Algorithm~\ref{alg:sew}.}

\noindent \textbf{Workflow Generation.} 
\sw{To generate workflows, we use an LLM to generate default workflows based on the given \sw{task} description $\mathcal{D}$~\footnote{Appendix~\ref{sec:prompt_corpus} shows all of the task descriptions.} and a template workflow $W^{temp}$}. 
\sw{A template workflow can be denoted with different workflow representation schemes. In particular, our \OURS{} explore five different schemes, namely \textit{Business Process Model and Notation (BPMN)}~\cite{white2004introduction}, \textit{Code Representation and Execution (CoRE)}~\cite{xu2024aios}, \textit{python}, \textit{YAML} and \textit{pseudo-code}, with their detailed description presented in Section~\ref{sec:rep}. Figure~\ref{fig:workflow_with_bpmn_and_core} shows two examples of the template workflow.}

From \sw{the workflow generation process as shown in Algorithm~\ref{alg:sew}}, we can obtain a set of default workflows $W^{def}$. \sw{Later, we will present how to use our workflow evolution module to rearrange and modify the structure of $W^{def}$.}

\noindent \textbf{Workflow Evolution.}
To formalise the workflow evolution process of \OURS{}, first we define a workflow $W$ represented with a certain representation scheme $rep$, where all $W$ in $rep$ are in textual format. 
\sw{We use the DE operator $\mathcal{F}(\cdot)$ to generate an evolved workflow as follows:}
\begin{equation}\label{eq:wf_evo}
    W' = \mathcal{F}(W_{def}
    |\mathcal{T}_\text{mut}),
\end{equation}
\sw{where $W'$ is the self-evolved workflow, $\mathcal{T}_\text{mut}$ is the mutation prompt and $\mathcal{F}(\cdot)$ representing the operation that an LLM takes $W_{def}$ and $\mathcal{T}_\text{mut}$ as input and output $W'$ (see Figure~\ref{fig:evoprompt} for more details)}.


It should be noted that the mutation prompt $\mathcal{T}_{mut}$ cannot ensure that $W'$ is a valid workflow. For example, $W'$ may not strictly follow the format of the \sw{representation scheme}. \sw{To measure the validity of $W'$, we define}
two \sw{rates}, namely the Logical Successful Rate (\textit{LSR}) and Generation Successful Rate (\textit{GSR}). \sw{The \textit{LSR} denotes the probability that generated $W'$ is valid, and the \textit{GSR} denotes the probability that the output of $W'$ is executable Python code.}
Specifically, 
$LSR = \frac{\sum_{i=1}^{|W'|} \mathbb{I}(\text{isValid}(W'_i))}{|W'|}$ and $GSR = \frac{\sum_{i=1}^{|W'|} \mathbb{I}(\text{isPython}(output(W'_i)))}{|W'|}$\;.
By measuring \sw{\textit{LSR} and \textit{GSR} of a certain representation scheme, we can determine which scheme is more suitable for \OURS{}.}

\noindent \textbf{Agent Evolution.}
After \sw{modifying the structure of workflows using the workflow evolution module}, the next step is to modify each agent's prompt. Similar to the workflow evolution, the agent evolution also relies on the mutation prompt. \sw{As mentioned earlier, we use} Direct Evolution (DE) and Hyper Evolution (HE) to improve an agent, where DE aims to modify an agent's prompt by directly applying a mutation prompt to it while HE first modifies the mutation prompt then apply the modified mutation prompt to \sw{an agent}. 

\begin{algorithm}[!t]
\begin{footnotesize}
\caption{Self-Evolving Workflow}
\label{alg:sew}
\KwIn{Task Description $\mathcal{D}$, Workflow Template $W^{temp}$, Mutation Prompt $\mathcal{T}_\text{mut}$, Hyper Mutation Prompt $\mathcal{T}_\text{hmut}$, Thinking-style Prompt $\mathcal{T}_\text{think}$}
\KwOut{Optimized Workflow $W’$}

\SetKwFunction{FMain}{SEW}
\SetKwProg{Fn}{Function}{:}{}
\Fn{\FMain{$\mathcal{T}_\text{des}, \mathcal{T}_\text{temp}, \mathcal{T}_\text{mut}, \mathcal{T}_\text{hmut}$, $\mathcal{T}_\text{think}$}}{
    \textbf{1. Workflow Generation}\;
    $W_{def} \gets \text{GenerateWorkflows}(\mathcal{T}_\text{des}, \mathcal{T}_\text{temp})$\;
    
    \textbf{2. Workflow Evolution}\;
    \For{each workflow $W_{def}$}{
        $W' \gets \mathcal{F}(W_{def}
    |\mathcal{T}_\text{mut})$\;
        
    }

    \textbf{3. Agent Evolution}\;
    \For{each agent $a$ in $W'$}{
        \textbf{3.1 Select Evolution Method}\;
        \If{First-order DE}{$a' \gets \mathcal{F}(a| \mathcal{T}_\text{mut})$\;}
        \ElseIf{Second-order DE}{$a'' \gets \mathcal{F}(\mathcal{F}(a| \mathcal{T}_\text{mut})|\mathcal{T}_\text{mut})$;}
        \ElseIf{Zero-order HE}{
            $a' \gets \mathcal{H}(a | \mathcal{H}(\mathcal{T}_\text{des} | \mathcal{T}_\text{think}))$\;}
        
        \ElseIf{First-order HE}{
            $a'' \gets \mathcal{H}(a|\mathcal{H}(\mathcal{T}_\text{mut}| \mathcal{T}_\text{hmut}))$\;}

    }
    
    \Return $W’$\;
}
\end{footnotesize}
\end{algorithm}

\noindent 
\textbf{Agentic Direct Evolution.}
To enhance the performance of an agent, \OURS{} directly apply the mutation prompt $\mathcal{T}_{mut}$ to an agent's prompt using the direct evolution operator as follows:
\begin{equation}\label{eq:directf}
    a' \gets \mathcal{F}(a| \mathcal{T}_\text{mut}),
\end{equation}
\sw{where $a$ is an agent and $a'$ is the agent with modified prompt, and we define the operation above as the first-order direct evolution.}

\sw{Based on the first-order direct evolution,} we propose the second-order direct evolution:
\begin{equation}\label{eq:directs}
    a'' \gets \mathcal{F}(\mathcal{F}(a| \mathcal{T}_\text{mut})|\mathcal{T}_\text{mut})
\end{equation}
By applying second-order direct evolution, we aim to further enhance the performance of an LLM agent.

\noindent \textbf{Agentic Hyper Evolution.}
Different from Direct Evolution, Hyper Evolution focuses on generating more effective mutation prompts. In other words, HE first modifies the mutation prompt $\mathcal{T}_{mut}$ then uses the new mutation prompt $\mathcal{T}_{mut}'$ to improve an agent's prompt. Formally, we define the zero-order hyper evolution as below:
\begin{equation}\label{eq:zhyper}
a' \gets \mathcal{H}(a | \mathcal{H}(\mathcal{T}_\text{des} | \mathcal{T}_\text{think}))
\end{equation}
where $\mathcal{T}_{think}$ are text descriptions of
general cognitive heuristics~\cite{promptbreeder}.

For zero-order HE, we use \sw{the} general cognitive heuristics $\mathcal{T}_{think}$ to generate useful prompts for solving problems described by the task description $\mathcal{D}$. Similar to how we use the mutation prompt $\mathcal{T}_{mut}$ to modify an agent's prompt, we can use a hyper-mutation prompt instead of $\mathcal{T}_{think}$ to modify $\mathcal{T}_{mut}$, which is defined as first-order HE.

From Eq~\ref{eq:zhyper}, a new mutation prompt is generated from the task description and some cognitive heuristics. In another way, we can use a hyper-mutation prompt to directly generate new variants from $\mathcal{T}_{mut}$ as follows:
\begin{equation}\label{eq:fhyper}
a'' \gets \mathcal{H}(a|\mathcal{H}(\mathcal{T}_\text{mut}| \mathcal{T}_\text{hmut}))
\end{equation}

Finally, by combining the workflow-evolution and agent evolution, our \OURS{} can generate more effective variants of workflows for solving the code generation task. In the next section, \sw{we will present and compare those five different representation schemes that can be leveraged by \OURS{}.}


\section{Workflow Representation}\label{sec:rep}
To generate a workflow using LLM, appropriate workflow textual representation schemes are essential. \sw{In fact, while it is straightforward to execute a workflow using code, representing it in natural language is non-trivial. A well-designed representation scheme should capture the structural and semantic components of a workflow and be easily interpreted by LLMs for downstream modification.}

As we discussed in the related work section, we explored five different textual representation schemes that can be used to denote workflows namely, \textit{Business Process Model and Notation (BPMN)}~\cite{white2004introduction}, \textit{Code Representation and Execution (CoRE)}~\cite{xu2024aios}, \textit{python}, \textit{YAML} and \textit{pseudo-code}, where each \sw{representation scheme} can be used to denote a workflow by text. \sw{The choice of these five schemes was driven by their distinct advantages in facilitating the representation and execution of agentic workflows, particularly in the context of self-evolving agentic workflows that our method, \OURS{}, aims to optimise.}

\begin{figure}[t]
    \centering
    \begin{subfigure}[b]{0.47\textwidth}
        \centering
        \includegraphics[width=\linewidth]{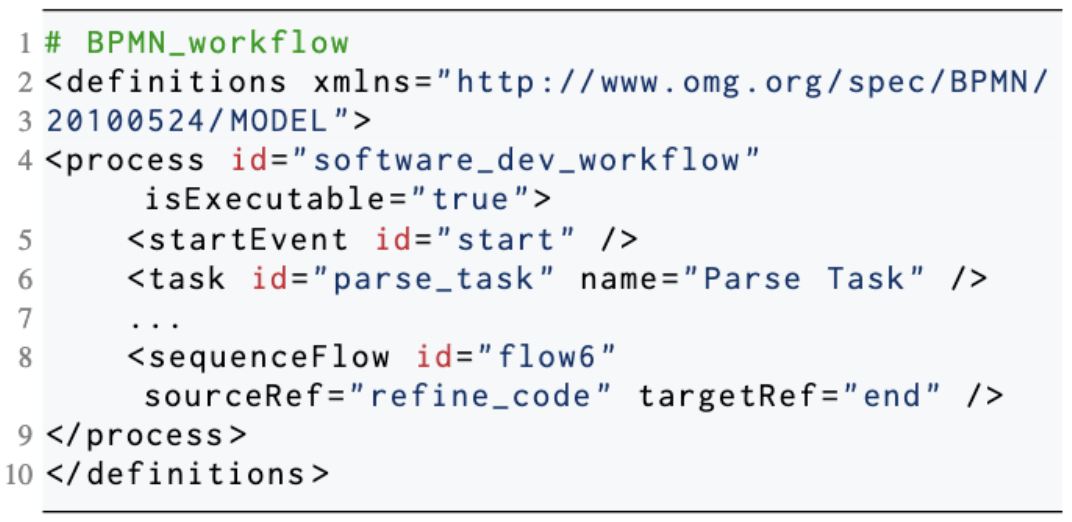}
        \label{fig:BPMN}
    \end{subfigure}
    \hfill
    \vspace{-1em}
    \begin{subfigure}[b]{0.47\textwidth}
        \centering
        \includegraphics[width=\linewidth]{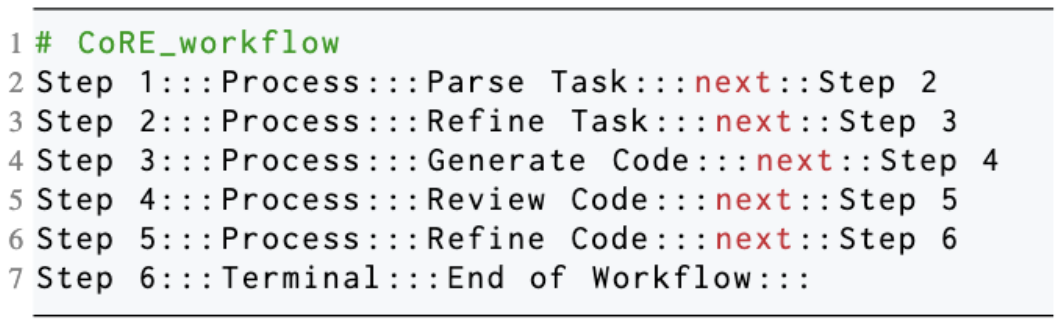}
        \label{fig:CoRE}
    \end{subfigure}
    \vspace{-0.7cm}
    \caption{A workflow represented by the BPMN and the CoRE schemes, respectively.}
    \label{fig:workflow_with_bpmn_and_core}
\end{figure}

\noindent  \textbf{BPMN}: This graphical standard is well-established in business process modeling and widely recognized for its ability to clearly depict the order of tasks and their dependencies.

\noindent  \textbf{CoRE}: CoRE integrates natural language programming, pseudo-code, and flow-based programming, and is a strong candidate for agentic workflows. It allows workflows to be directly executable and interpretable by LLMs, offering advantages for our self-evolving framework.

\noindent  \textbf{Python}: As a widely adopted programming language, Python is not only familiar to many practitioners but also flexible in terms of representing workflows through its readable syntax and extensive ecosystem of libraries. For agentic workflows requiring programmatic execution, Python allows for easy integration and adaptation of agents into working solutions.

\noindent  \textbf{YAML}: YAML is a human-readable data serialisation format widely used for configuration files and workflow definitions due to its simplicity and readability. YAML’s flexibility in representing hierarchical data structures makes it well-suited for workflows that need to be configured or defined by humans but executed by machines. 

\noindent  \textbf{Pseudo-code}: Pseudo-code is a high-level representation that is often used for illustrating algorithms and workflows in a way that is easy for both humans and machines to understand. Pseudo-code offers an abstraction that bridges natural language and formal code, making it an excellent choice for expressing workflows that need to be easily read and modified. 

To clearly illustrate the differences between workflow representation schemes, we present an example agentic workflow represented using both the BPMN and CoRE schemes in Figure \ref{fig:workflow_with_bpmn_and_core}. In Figure \ref{fig:workflow_with_bpmn_and_core}, a software development pipeline, consisting of sequential tasks such as parsing input, refining content, generating code, reviewing, and iterating improvements, is represented by BPMN and CoRE, respectively. Each stage is represented as a task node, while dependencies between tasks are captured as sequence flows, ensuring clear process execution. Although denoted with different \sw{representation schemes}, they shall perform the same function when executed\footnote{Additional representation formats, such as Python, pseudo-code and YAML, are provided in Appendix~\ref{sec:workflowRM}}.

\sw{These five schemes were chosen for their diverse capabilities in representing workflows and their practical utility in a self-evolving framework, where agents and workflows are dynamically generated and optimised. Our exploration of these schemes aims to identify the most suitable representation for evolving agentic workflows in code generation tasks, where LLMs are leveraged for both understanding and executing the workflows.}

\section{Experiments}\label{sec:exp}
\subsection{Dataset}
To examine our proposed \OURS{} framework, we choose the
LiveCodeBench (LCB)~\cite{jain2024livecodebench} dataset, which is a comprehensive benchmark designed to evaluate the coding capabilities of LLMs. 
We randomly sampled 100 samples from the code generation subset of LCB\footnote{\sw{We use the release\_v1 version of the code generation subset of LCB.}} for validation and the remaining 300 samples for testing. \sw{In addition, we also use the MBPP~\cite{mbpp} and HumanEval+~\cite{humaneval} datasets following the data split in AFlow~\cite{zhang2024aflow}. }
To evaluate performance on the code generation task, each method is required to generate 10 candidate solutions per sample. We use pass@1, pass@5, and pass@10 as evaluation metrics.

\subsection{Baselines}
\sw{We compare our proposed SEW against five baseline prompting techniques across two different backbone models (i.e. Claude Opus 4.6 and GPT-5.3) on three code generation tasks (i.e HumanEval+, MBPP, and LCB): (1)	Backbone Models (Claude Opus 4.6 and GPT-5.3). (2) Chain-of-Thought (CoT)~\cite{cot} uses reasoning steps explicitly stated within the prompt. (3) Automated Design of Agentic Systems (ADAS)~\cite{adas}: A methodology that leverages meta-agent frameworks to automatically design and optimise agentic systems. (4) AFlow~\cite{zhang2024aflow}: An automated framework that efficiently explores and optimises agentic workflows using Monte Carlo Tree Search. (5) PromptBreeder~\cite{promptbreeder} is a gradient-free evolutionary framework that improves agents by iteratively mutating and selecting prompt variants.}

\begin{table}[]
    \centering
    \small
    \resizebox{\linewidth}{!}{
    \begin{tabular}{lccc}
    \toprule
         Method& HumanEval+& MBPP& LCB \\
         \midrule
         \rowcolor{gray!40}\multicolumn{4}{c}{\textbf{Claude Opus 4.6}}\\
         Claude Opus 4.6& 91.5 & 85.3 &52.0\\
        CoT & 93.2& 87.5& 53.5\\
        \sw{PromptBreeder} & 94.0& 89.0& 55.0\\
         ADAS & 93.6& 88.0&54.2\\
         AFlow & 94.3& 89.5& -\\
         \midrule
         \OURS{} (Claude Opus 4.6)& 94.8& \textbf{91.2}& \textbf{57.2}\\
         \midrule
         \rowcolor{gray!40}\multicolumn{4}{c}{\textbf{GPT-5.3}}\\
         GPT-5.3& 93.1& 79.4& 50.0\\
        CoT & 94.5& 82.0& 51.5\\
         \sw{PromptBreeder} & 95.2& 84.5& 53.0\\
         ADAS & 94.8& 83.0& 52.2\\
         AFlow & 95.5& 85.0& - \\
         \midrule
         \OURS{} (GPT-5.3) & \textbf{95.6}& 86.7& 55.5\\
         \bottomrule
    \end{tabular}}
    \caption{Performance comparison (pass@1) between our \OURS{} and baselines. ‘-’ refers to out-of-time errors, where the LLM executor has been trapped in executing accidental scripts with infinite loops. We adopt two LLMs, i.e., Claude Opus 4.6 and GPT-5.3, as backbone models for all methods.}
    \label{tab:overall}
    \vspace{-0.3cm}
\end{table}
\subsection{Experimental Setup}
\sw{We conduct an exhaustive search on self-evolved workflows $W'$, represented by the following methods, including BPMN, CoRE, python, YAML and pseudo-code. We use all mutation prompts to evolve workflows represented by 5 schemes. 	Although various types of workflows are generated during the self-evolution process, not all of them are valid for code generation tasks. Among all generated workflows, the \textbf{task parsing workflow} and \textbf{code rewriting workflow}\footnote{We also report the token cost of these two workflows in the Appendix~\ref{sec:token}} are more effective than the other counterparts. In particular, variants based on these two workflows can largely outperform competitive baselines, hence, we choose the best variant to represent our \OURS{}.}.

\begin{table}
    \centering
    \large
    \resizebox{0.48\textwidth}{!}{
    \begin{tabular}{lccccc}
        \toprule
        Rep method & BPMN & CoRE & python & YAML & pseudo-code \\
        \midrule
        \textit{LSR} & \textbf{91.5\%} & 78.5\% & 91.0\%& 83.5\% & 60.0\%\\
        \textit{GSR} & 58.0\% & \textbf{80.5\%} & 50.0\% & 60.5\% & 45.0\% \\
        \bottomrule
    \end{tabular}}
    \caption{Logic Successful Rate (\textit{LSR}) and Generation Successful Rate (\textit{GSR}) for Business Process Model and Notation, Code Representation and Execution, python, YAML and pseudo-code.}
    \vspace{-0.2cm}
    \label{tab:srate}
\end{table}

\subsection{Main Results}
\sw{To compare the performance of SEW and other baselines, we adopt two backbone models i.e., Claude Opus 4.6 and GPT-5.3. From Table~\ref{tab:overall}, we find that (1) \OURS{} can largely outperform those two backbone models at both settings; (2) \OURS{} is more effective than CoT, a robust prompting technique for enhancing LLM's ability to solve complex tasks by breaking them down into sequential thought processes; (3) when leveraging the same backbone model, our \OURS{} outperform other state-of-the-art workflow designing methods such as ADAS and AFlow. Therefore, we can conclude that our \OURS{} framework is more effective than different types of baselines under the same setting in the code generation task. 	In addition, we observe that across the three datasets, methods using Claude Opus 4.6 as the backbone generally outperform those using GPT-5.3 (on both MBPP and LCB). Hence, to save space, we report only the analysis of \OURS{} (Claude Opus 4.6) in the following sections.}

\subsection{Analysis}
\tcbset{enhanced,
  boxrule=0pt, frame hidden,
  borderline west={4pt}{0pt}{gray!75!black},
  top=0pt, bottom=0pt,
  colback=gray!10!white,
  sharp corners}
\begin{tcolorbox}
\textbf{RQ1}: Which \sw{scheme} is the most effective for structuring agentic workflows?
\end{tcolorbox}

To identify the most suitable workflow \sw{scheme} for LLMs among the five, we conducted an exhaustive search using various mutation prompts. For a given workflow  $W$  represented in Python, 100 different mutation prompts generated 100 variants. If 50 of these variants are parsable and 30 can generate executable codes, the \textit{LSR} and \textit{GSR} for Python are 50\% and 30\%, respectively. Notably, \textit{LSR} is always greater than or equal to \textit{GSR}, as not all parsed workflows can complete the task.

As shown in Table~\ref{tab:srate}, BPMN achieved the highest LSR at 91.5\%. However, their GSR performance was suboptimal, whereas the recently proposed CoRE method achieved the best GSR. This suggests that while traditional BPMN and Python representations are easier for LLMs to parse, the CoRE method -- which integrates natural language programming, pseudo-code programming, and flow programming -- is the most effective for workflow representation. We therefore conclude that CoRE enables optimal comprehension and utilisation when denoting agentic workflows.

\begin{table}[]
    \centering
    \small
        \resizebox{\linewidth}{!}{

    \begin{tabular}{lccc}
    \toprule
         Method& HumanEval+& MBPP& LCB \\
         \midrule
        Claude Opus 4.6 & 91.5 & 85.3 & 52.0\\
        \midrule
         \multicolumn{4}{c}{\textbf{task parsing workflow}}\\
         \hline
        \sw{Only Workflow-Evolved}& 93.2& 88.8& 55.0\\
         \sw{Workflow + Agent Co-Evolved} & 94.6& 93.0& 60.2\\
         \%improv.& 1.5& 4.7& 9.5\\
         \midrule
         \multicolumn{4}{c}{\textbf{code rewriting workflow}}\\
         \hline
         \sw{Only Workflow-Evolved}& 94.0& 87.2& 54.0\\
         \sw{Workflow + Agent Co-Evolved} & 95.2 & 91.0& 58.5\\
         \%improv.& 1.3 & 4.4& 8.3\\
         \bottomrule
    \end{tabular}}
    \caption{Performance comparison (pass@1) between the default version of two representative workflows generated from workflow evolution and their improved variants using agent evolution. All workflows use Claude Opus 4.6 as their backbone model.}
    \label{tab:rq3}
    \vspace{-0.3cm}
\end{table}

\tcbset{enhanced,
  boxrule=0pt, frame hidden,
  borderline west={4pt}{0pt}{gray!75!black},
  top=0pt, bottom=0pt,
  colback=gray!10!white,
  sharp corners}
\begin{tcolorbox}
\textbf{RQ 2}: How do SEW's workflow evolution and agent evolution modules affect the performance of coding generation?
\end{tcolorbox}

\begin{figure*}[t]
    \centering
    \includegraphics[width=\linewidth]{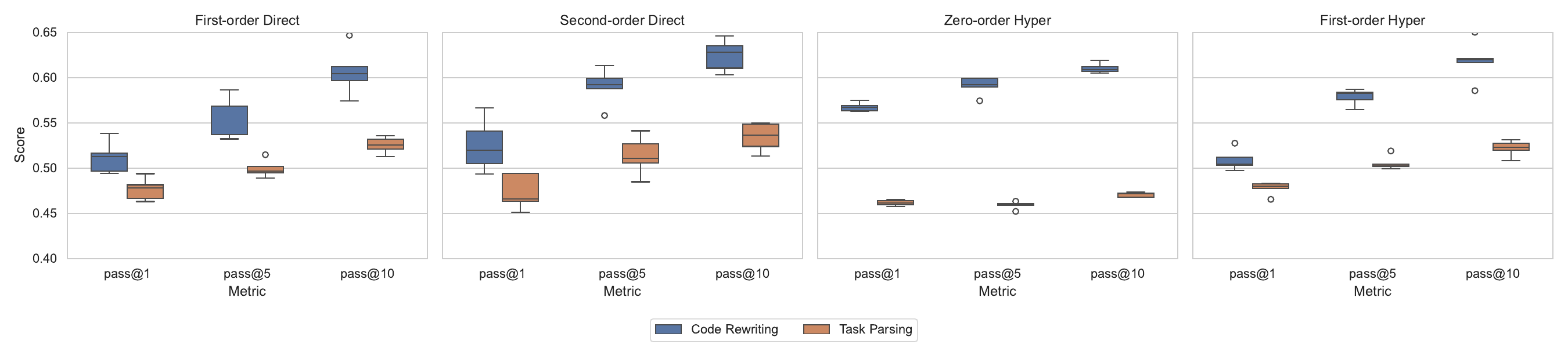}
    \vspace{-2mm}
    \caption{Performance comparison of Code Rewriting and Task Parsing Workflows under different agent evolution strategies on the LCB dataset.}
    \vspace{-0.4cm}
    \label{fig:comp}
\end{figure*}

\sw{To understand how our workflow evolution and
agent evolution modules affect the performance of workflows generated by \OURS{}, we select two representative workflows generated by \OURS{}, namely \textbf{task parsing workflow} and \textbf{code rewriting workflow}. We chose these two workflows since most of the variants built upon these two workflows can bring large improvements. Specifically, the task parsing workflow leverages an agent to first parse the task and then send the parsed result to a coding agent to generate the code subsequently. In comparison, a code rewriting workflow incorporates a code generation agent to generate the initial outcome and then uses the code reviewing agent to determine if this outcome can pass the test followed by a code rewriting agent to rewrite the code based on the feedback from the code reviewing agent.\footnote{Details of these two workflows can be found in Appendix~\ref{sec:prompt_corpus}.} }

\sw{Notably, the workflow evolution module is designed to generate novel workflow structures, while the agent evolution module focuses on creating effective prompts for each agent.
In particular, we compare: (1) workflows generated by the workflow evolution module versus those produced by the backbone model, and (2) workflows generated by the workflow evolution module versus those that incorporate both workflow and agent evolutions. As shown in Table~\ref{tab:rq3}, the task parsing and code rewriting workflows produced by \OURS{} consistently outperform the Claude Opus 4.6 backbone model across three datasets. This initial improvement suggests that our workflow evolution module generates novel workflow topologies more effectively than relying solely on the LLM. Building on these novel workflows, the agent evolution module further enhances performance by generating high-quality prompts for each agent. Specifically, our agent evolution module improves the performance of the task parsing workflow by 9.5\% on the LCB dataset. In summary, our results demonstrate that the workflow evolution module effectively produces novel workflow structures, and the agent evolution module further unlocks their potential by injecting high-quality prompts.}

\tcbset{enhanced,
  boxrule=0pt, frame hidden,
  borderline west={4pt}{0pt}{gray!75!black},
  top=0pt, bottom=0pt,
  colback=gray!10!white,
  sharp corners}
\begin{tcolorbox}
\textbf{RQ 3}: 
How do different agentic evolution strategies affect the performance of workflows generated by SEW?
\end{tcolorbox}
We have introduced the \textbf{Direct Evolution} (DE) and \textbf{Hyper Evolution} (HE) operators, where for each we proposed its corresponding lower-order and higher-order versions. To examine the effectiveness of different operators, we randomly sampled five different mutation prompts and used these randomly sampled mutation prompts to generate five different variants for both workflows mentioned earlier for each operator. We use four box plots to illustrate the performance distribution of these two workflows on the LCB dataset.

\sw{From Figure~\ref{fig:comp}, we can observe that HE consistently demonstrates lower variance than DE by comparing the first row and second row of Figure~\ref{fig:comp}. The variance of both workflows under the zero-order hyper evolution is especially small. This indicates that the HE operator, particularly zero-order HE, exhibit superior robustness compared to DE, as they are less sensitive to variations in mutation prompts across different tasks. In terms of best performance, DE, especially second-order DE, tends to achieve higher peak performance in certain metrics, such as pass@10 for Code Rewriting Workflow, where it reaches up to around 0.57–0.60. This suggests that DE can optimize for specific high-performance outcomes. On the other hand, HE, while slightly lower in peak performance, provides a more balanced and reliable performance profile, making it more suitable for consistency.}

\sw{Therefore, the choice between DE and HE depends on the requirements of the task: DE is preferable for maximizing performance, while HE is better suited for real-world applications where robustness is more important. In addition, higher-order evolutions (Second-order DE and First-order HE) are better suited for tasks that require maximizing performance and can tolerate some variability, while lower-order evolutions (First-order DE and Zero-order HE) provide higher robustness.}

\section{Conclusion}
In this paper, we introduce Self-Evolving Workflow (\OURS{}), a novel framework that enables LLM-based multi-agent workflows to automatically adapt and evolve for improved performance in automatic code generation. Unlike conventional hand-crafted workflows, \OURS{} leverages self-evolving prompting to optimise both workflow structures and individual agent capabilities. Through extensive experiments on three coding generation benchmarks, we demonstrate that self-evolved workflows consistently outperform single-agent baselines, even when the latter are enhanced with improved prompts. Moreover, our results show that higher-order evolution strategies further improve workflow effectiveness by refining task execution dynamics. Among different workflow \sw{representation schemes}, CoRE emerges as the most effective, offering a balance between logical correctness and execution success. These findings suggest that \OURS{} provides a promising direction for agentic workflow optimisation, reducing reliance on manual workflow design and prompt engineering while improving adaptability and efficiency.

\section*{Limitations}
While \OURS{} proves effective for code generation, its generalization to other AI-driven tasks, such as reasoning or planning, remains unknown. Another limitation lies in the workflow execution constraints, where some generated workflows, despite being logically sound, fail to produce executable outputs, suggesting a need for more robust validation mechanisms. Furthermore, the effectiveness of \OURS{} is dependent on the capabilities of the underlying LLM, meaning that future advancements in LLM architectures could significantly impact its performance. 
Addressing these limitations in future work will be crucial for extending \OURS{}’s applicability and enhancing its adaptability for broader tasks.
\bibliography{references}

@article{huang2023agentcoder,
  title={Agentcoder: Multi-agent-based code generation with iterative testing and optimisation},
  author={Huang, Dong and Zhang, Jie M and Luck, Michael and Bu, Qingwen and Qing, Yuhao and Cui, Heming},
  journal={arXiv preprint arXiv:2312.13010},
  year={2023}
}

@inproceedings{islam2024mapcoder,
  title={MapCoder: Multi-Agent Code Generation for Competitive Problem Solving},
  author={Islam, Md Ashraful and Ali, Mohammed Eunus and Parvez, Md Rizwan},
  booktitle={Proceedings of the 62nd Annual Meeting of the Association for Computational Linguistics (Volume 1: Long Papers)},
  pages={4912--4944},
  year={2024}
}

@article{mbpp,
  title={Program synthesis with large language models},
  author={Austin, Jacob and Odena, Augustus and Nye, Maxwell and Bosma, Maarten and Michalewski, Henryk and Dohan, David and Jiang, Ellen and Cai, Carrie and Terry, Michael and Le, Quoc and others},
  journal={arXiv preprint arXiv:2108.07732},
  year={2021}
}

@article{opsahl2024optimizing,
  title={Optimizing instructions and demonstrations for multi-stage language model programs},
  author={Opsahl-Ong, Krista and Ryan, Michael J and Purtell, Josh and Broman, David and Potts, Christopher and Zaharia, Matei and Khattab, Omar},
  journal={arXiv preprint arXiv:2406.11695},
  year={2024}
}

@article{adas,
  title={Automated design of agentic systems},
  author={Hu, Shengran and Lu, Cong and Clune, Jeff},
  journal={arXiv preprint arXiv:2408.08435},
  year={2024}
}

@article{cot,
  title={Chain-of-thought prompting elicits reasoning in large language models},
  author={Wei, Jason and Wang, Xuezhi and Schuurmans, Dale and Bosma, Maarten and Xia, Fei and Chi, Ed and Le, Quoc V and Zhou, Denny and others},
  journal={Advances in neural information processing systems},
  volume={35},
  pages={24824--24837},
  year={2022}
}

@article{liu2024your,
  title={Is your code generated by chatgpt really correct? rigorous evaluation of large language models for code generation},
  author={Liu, Jiawei and Xia, Chunqiu Steven and Wang, Yuyao and Zhang, Lingming},
  journal={Advances in Neural Information Processing Systems},
  volume={36},
  year={2024}
}

@article{qian2023communicative,
  title={Communicative agents for software development},
  author={Qian, Chen and Cong, Xin and Yang, Cheng and Chen, Weize and Su, Yusheng and Xu, Juyuan and Liu, Zhiyuan and Sun, Maosong},
  journal={arXiv preprint arXiv:2307.07924},
  volume={6},
  number={3},
  year={2023}
}

@article{zhang2025evoflow,
  title={EvoFlow: Evolving Diverse Agentic Workflows On The Fly},
  author={Zhang, Guibin and Chen, Kaijie and Wan, Guancheng and Chang, Heng and Cheng, Hong and Wang, Kun and Hu, Shuyue and Bai, Lei},
  journal={arXiv preprint arXiv:2502.07373},
  year={2025}
}

@inproceedings{liu2024large,
  title={Large language models as evolutionary optimizers},
  author={Liu, Shengcai and Chen, Caishun and Qu, Xinghua and Tang, Ke and Ong, Yew-Soon},
  booktitle={2024 IEEE Congress on Evolutionary Computation (CEC)},
  pages={1--8},
  year={2024},
  organization={IEEE}
}

@article{agarwal2024promptwizard,
  title={PromptWizard: Task-Aware Agent-driven Prompt Optimization Framework},
  author={Agarwal, Eshaan and Dani, Vivek and Ganu, Tanuja and Nambi, Akshay},
  journal={arXiv preprint arXiv:2405.18369},
  year={2024}
}

@article{zhou2025multi,
  title={Multi-Agent Design: Optimizing Agents with Better Prompts and Topologies},
  author={Zhou, Han and Wan, Xingchen and Sun, Ruoxi and Palangi, Hamid and Iqbal, Shariq and Vuli{\'c}, Ivan and Korhonen, Anna and Ar{\i}k, Sercan {\"O}},
  journal={arXiv preprint arXiv:2502.02533},
  year={2025}
}

@article{zhou2024symbolic,
  title={Symbolic learning enables self-evolving agents},
  author={Zhou, Wangchunshu and Ou, Yixin and Ding, Shengwei and Li, Long and Wu, Jialong and Wang, Tiannan and Chen, Jiamin and Wang, Shuai and Xu, Xiaohua and Zhang, Ningyu and others},
  journal={arXiv preprint arXiv:2406.18532},
  year={2024}
}

@article{zhang2024aflow,
  title={Aflow: Automating agentic workflow generation},
  author={Zhang, Jiayi and Xiang, Jinyu and Yu, Zhaoyang and Teng, Fengwei and Chen, Xionghui and Chen, Jiaqi and Zhuge, Mingchen and Cheng, Xin and Hong, Sirui and Wang, Jinlin and others},
  journal={arXiv preprint arXiv:2410.10762},
  year={2024}
}

@article{humaneval,
  title={Evaluating large language models trained on code},
  author={Chen, Mark and Tworek, Jerry and Jun, Heewoo and Yuan, Qiming and Pinto, Henrique Ponde De Oliveira and Kaplan, Jared and Edwards, Harri and Burda, Yuri and Joseph, Nicholas and Brockman, Greg and others},
  journal={arXiv preprint arXiv:2107.03374},
  year={2021}
}

@inproceedings{promptbreeder,
  title={Promptbreeder: Self-Referential Self-Improvement via Prompt Evolution},
  author={Fernando, Chrisantha and Banarse, Dylan Sunil and Michalewski, Henryk and Osindero, Simon and Rockt{\"a}schel, Tim},
  booktitle={International Conference on Machine Learning},
  pages={13481--13544},
  year={2024},
  organization={PMLR}
}

@article{jain2024livecodebench,
  title={Livecodebench: Holistic and contamination free evaluation of large language models for code},
  author={Jain, Naman and Han, King and Gu, Alex and Li, Wen-Ding and Yan, Fanjia and Zhang, Tianjun and Wang, Sida and Solar-Lezama, Armando and Sen, Koushik and Stoica, Ion},
  journal={arXiv preprint arXiv:2403.07974},
  year={2024}
}

@article{xu2024aios,
  title={AIOS Compiler: LLM as Interpreter for Natural Language Programming and Flow Programming of AI Agents},
  author={Xu, Shuyuan and Li, Zelong and Mei, Kai and Zhang, Yongfeng},
  journal={CoRR},
  year={2024}
}

@article{white2004introduction,
  title={Introduction to BPMN},
  author={White, Stephen A},
  journal={Ibm Cooperation},
  volume={2},
  number={0},
  pages={0},
  year={2004}
}

@article{chi2024sela,
  title={SELA: Tree-Search Enhanced LLM Agents for Automated Machine Learning},
  author={Chi, Yizhou and Lin, Yizhang and Hong, Sirui and Pan, Duyi and Fei, Yaying and Mei, Guanghao and Liu, Bangbang and Pang, Tianqi and Kwok, Jacky and Zhang, Ceyao and others},
  journal={arXiv preprint arXiv:2410.17238},
  year={2024}
}

@article{yuan2024evoagent,
  title={Evoagent: Towards automatic multi-agent generation via evolutionary algorithms},
  author={Yuan, Siyu and Song, Kaitao and Chen, Jiangjie and Tan, Xu and Li, Dongsheng and Yang, Deqing},
  journal={arXiv preprint arXiv:2406.14228},
  year={2024}
}

@article{gao2024agentscope,
  title={Agentscope: A flexible yet robust multi-agent platform},
  author={Gao, Dawei and Li, Zitao and Pan, Xuchen and Kuang, Weirui and Ma, Zhijian and Qian, Bingchen and Wei, Fei and Zhang, Wenhao and Xie, Yuexiang and Chen, Daoyuan and others},
  journal={arXiv preprint arXiv:2402.14034},
  year={2024}
}

@article{hong2023metagpt,
  title={Metagpt: Meta programming for multi-agent collaborative framework},
  author={Hong, Sirui and Zheng, Xiawu and Chen, Jonathan and Cheng, Yuheng and Wang, Jinlin and Zhang, Ceyao and Wang, Zili and Yau, Steven Ka Shing and Lin, Zijuan and Zhou, Liyang and others},
  journal={arXiv preprint arXiv:2308.00352},
  year={2023}
}

@article{zhang2024pybench,
  title={PyBench: Evaluating LLM Agent on various real-world coding tasks},
  author={Zhang, Yaolun and Pan, Yinxu and Wang, Yudong and Cai, Jie},
  journal={arXiv preprint arXiv:2407.16732},
  year={2024}
}

@article{xu2024theagentcompany,
  title={Theagentcompany: benchmarking llm agents on consequential real world tasks},
  author={Xu, Frank F and Song, Yufan and Li, Boxuan and Tang, Yuxuan and Jain, Kritanjali and Bao, Mengxue and Wang, Zora Z and Zhou, Xuhui and Guo, Zhitong and Cao, Murong and others},
  journal={arXiv preprint arXiv:2412.14161},
  year={2024}
}

@article{qiao2023taskweaver,
  title={Taskweaver: A code-first agent framework},
  author={Qiao, Bo and Li, Liqun and Zhang, Xu and He, Shilin and Kang, Yu and Zhang, Chaoyun and Yang, Fangkai and Dong, Hang and Zhang, Jue and Wang, Lu and others},
  journal={arXiv preprint arXiv:2311.17541},
  year={2023}
}

@article{xiao2024flowbench,
  title={FlowBench: Revisiting and Benchmarking Workflow-Guided Planning for LLM-based Agents},
  author={Xiao, Ruixuan and Ma, Wentao and Wang, Ke and Wu, Yuchuan and Zhao, Junbo and Wang, Haobo and Huang, Fei and Li, Yongbin},
  journal={CoRR},
  year={2024}
}

@article{li2025structured,
  title={Structured chain-of-thought prompting for code generation},
  author={Li, Jia and Li, Ge and Li, Yongmin and Jin, Zhi},
  journal={ACM Transactions on Software Engineering and Methodology},
  volume={34},
  number={2},
  pages={1--23},
  year={2025},
  publisher={ACM New York, NY}
}

@inproceedings{zhang2024effectiveness,
  title={On the effectiveness of large language models for github workflows},
  author={Zhang, Xinyu and Muralee, Siddharth and Cherupattamoolayil, Sourag and Machiry, Aravind},
  booktitle={Proceedings of the 19th International Conference on Availability, Reliability and Security},
  pages={1--14},
  year={2024}
}

@inproceedings{zhoularge,
  title={Large language models are human-level prompt engineers},
  author={Zhou, Yongchao and Muresanu, Andrei Ioan and Han, Ziwen and Paster, Keiran and Pitis, Silviu and Chan, Harris and Ba, Jimmy},
  booktitle={The Eleventh International Conference on Learning Representations},
  year={2022}
}
\appendix

\section{Appendix}
\label{sec:appendix}
\subsection{Workflow \sw{Representation Schemes}}\label{sec:workflowRM}
\sw{In Section~\ref{sec:rep}, we have introduced five different representation schemes namely, \textit{Business Process Model and Notation (BPMN)}~\cite{white2004introduction}, \textit{Code Representation and Execution (CoRE)}~\cite{xu2024aios}, \textit{python}, \textit{YAML} and \textit{pseudo-code}, where we used two scripts (Figure~\ref{fig:workflow_with_bpmn_and_core}) to denote a workflow represented by BPMN and CoRE, respectively. Below, we will show the same workflow represented by python, YAML and pseudo-code.}

\begin{samepage}\label{prg1}
\begin{lstlisting}[language=Python,breaklines=true,showstringspaces=false]
# python_workflow
steps = [
    {'name': 'task_parsing_agent', 'args': ['task_description'], 'output': 'parsed_task'},
    {'name': 'task_refinement_agent', 'args': ['task_description', 'parsed_task'], 'output': 'refined_task'},
    {'name': 'code_generation_agent', 'args': ['refined_task'], 'output': 'generated_code'},
    {'name': 'code_reviewer_agent', 'args': ['refined_task', 'generated_code'], 'output': 'review_comments'},
    {'name': 'code_refinement_agent', 'args': ['refined_task', 'review_comments'], 'output': 'refined_code'},
]
\end{lstlisting}
\end{samepage}

\begin{samepage}
\begin{lstlisting}[language=Python,breaklines=true,showstringspaces=false]
# pseudo_code_workflow
task_parsing_agent(task_description) -> parsed_task
task_refinement_agent(task_description, parsed_task) -> refined_task
code_generation_agent(refined_task) -> generated_code
code_reviewer_agent(refined_task, generated_code) -> review_comments
code_refinement_agent(refined_task, review_comments) -> refined_code
\end{lstlisting}
\end{samepage}

\begin{samepage}
\begin{lstlisting}[language=Python,breaklines=true,showstringspaces=false]
# YAML_workflow
- name: task_parsing_agent
  args:
    - task_description
  output: parsed_task

- name: task_refinement_agent
  args:
    - task_description
    - parsed_task
  output: refined_task

- name: code_generation_agent
  args:
    - refined_task
  output: generated_code

- name: code_reviewer_agent
  args:
    - refined_task
    - generated_code
  output: review_comments

- name: code_refinement_agent
  args:
    - refined_task
    - review_comments
  output: refined_code
\end{lstlisting}
\end{samepage}

\subsection{Prompt Corpus}\label{sec:prompt_corpus}
\sw{In Section~\ref{sec:sew}, we have introduced many textual variables including the task description, generation prompts and evolutionary prompts. In this section, we will detail the textual information.}

\begin{tcolorbox}
\textcolor{red}{Task Description $\mathcal{D}$ for each dataset:} 

\textbf{LiveCodeBench}: The code generation task in LiveCodeBench involves generating correct and functional code from a natural language problem description, where the model is evaluated based on its ability to pass a set of unseen test cases.\\

\textbf{HumanEval+}: The HumanEval+ dataset, developed by OpenAI, comprises 164 handcrafted programming problems, each including a function signature, docstring, body, and multiple unit tests, designed to evaluate the code generation capabilities of large language models by assessing their ability to generate functionally correct code from docstrings.\\

\textbf{MBPP}: The MBPP (Mostly Basic Python Problems) dataset comprises approximately 1,000 crowd-sourced Python programming problems, each including a task description, code solution, and three automated test cases, designed to be solvable by entry-level programmers and covering programming fundamentals and standard library functionality.
\end{tcolorbox}

We use $W^{temp}$ to denote a template workflow, which is used to guide LLMs to generate default workflows. In particular, we use the example workflows listed in Appendix~\ref{sec:workflowRM} as our templates. 

\begin{tcolorbox}
\textcolor{red}{Default agent generation prompt:}

You are an AI prompt engineer. Your task is to create specific prompts for each agent in the provided workflow. Please follow these steps:
\begin{enumerate}
	\item Understand the Workflow: Here is the detailed workflow: [Detailed workflow]
    \item Identify Agent Roles: Based on the workflow, determine the distinct roles and responsibilities of each agent involved.
    \item Generate Agent-Specific Prompts: For each identified agent, craft a clear and concise prompt that includes:
    \begin{itemize}
    \item Agent Role: A brief description of the agent’s function within the workflow.
    \item Objectives: The specific goals the agent is expected to achieve.
    \item Inputs: The information or data the agent will receive.
    \item  Outputs: The expected results or actions the agent should produce.
    \end{itemize}
\end{enumerate}

\textcolor{red}{Default workflow generation prompt:}

You are an AI workflow designer. Your task is to create a detailed Agent Workflow tailored to the provided workflow template and dataset description. Please follow these steps:
\begin{enumerate}
    \item Review the Workflow Template: [Detailed workflow template]
    \item Analyze the Dataset Description: [Dataset description]
    \item Design the Agent Workflow: Based on the above information, develop a comprehensive Agent Workflow that includes:
    \begin{itemize}
        \item Inputs and Outputs: Define the types of input data and the expected output results.
        \item Steps and Sequence: Outline each step of the workflow and specify the order of execution.
        \item Agent Roles and Responsibilities: Describe the role and duties of the agent at each step.
    \end{itemize}
\end{enumerate}
\end{tcolorbox}

Our used evolutionary prompts, including the \textit{mutation prompt}, \textit{hyper-mutation prompt} and \textit{thinking-style} prompt, can be found in the appendix of PromptBreeder~\cite{promptbreeder}.

\sw{In Table~\ref{tab:rq3}, we have presented two workflows obtained from \OURS{} namely the Task Parsing Workflow and Code Rewriting Workflow. In the following, we will show the name of each agent and its corresponding prompt in those workflows. In particular, we will first show the version that has not been improved by the agent evolution module, followed by the version that has been improved by the agent evolution module.}

\footnotetext{Notably, workflows may contain loops, for example, a loop of Code Reviewer Agent and Code Rewriting Agent. However, we will not repeat the prompt for the same agent repetitively.}

\subsubsection{Workflows without agent evolution module}
\begin{tcolorbox}[colback=gray!5!white, colframe=gray!75!black, breakable,]
\textcolor{red}{Task Parsing Workflow:}

\textcolor{blue}{1. Task Parsing Agent:} "You are a task parsing agent. Comprehensively summarize the given programming task for the subsequent code generation. You will NOT return anything except for the task summary. \{ \textcolor{cyan}{TASK PROMPT}\}

\textcolor{blue}{2. Code Generation Agent:} 
\detokenize{
"You are a proficient Python programmer. Your task is to write Python code according to the summary parsed by your colleague. You will be given the problem description followed by the summary. You will NOT return anything except for the program."
}$\oplus$ \{ \textcolor{cyan}{Output from Task Parsing Agent}\}

\textcolor{red}{Code Rewriting Workflow:}

\textcolor{blue}{1. Code Generation Agent:}
\detokenize{"You are a proficient Python programmer. Your task is to write Python code according to the summary parsed by your colleague. You will be given the problem description followed by the summary. You will NOT return anything except for the program."
}$\oplus$ \{ \textcolor{cyan}{TASK PROMPT}\}

\textcolor{blue}{2. Code Reviewer Agent:} \detokenize{     "You are a critical python code reviewer. You are tasked to label generated codes with 1 or 0, where 1 indicates that this code satisfies the requirements and can pass the sample test, 
while 0 indicates that this code doesn't satisfies the requirements and will fail the sample test. You will be given the Problem Description followed by the corresponding Generated Code. You will NOT return anything except for the numerical label."
} $\oplus$ \{ \textcolor{cyan}{Output from Code Generation Agent}\}

\textcolor{blue}{3. Code Rewriting Agent: }\detokenize{"You are a proficient Python programmer tasked with coding solutions based on given problem specifications. You just generated some codes that cannot pass the sample test. 
You role is to regenerate python code that strictly adheres to the specifications, ensuring it reads input from standard input (stdin) and writes output to standard output (stdout). 
You will be given the Problem Description followed by the Comments and Reasons why your previous code fails. You will NOT return anything except for the program."
} $\oplus$ \{ \textcolor{cyan}{Output from Code Reviewer Agent}\}
\end{tcolorbox}

\subsubsection{Workflows improved by agent evolution module}
\begin{tcolorbox}[colback=gray!5!white, colframe=gray!75!black,breakable,]
\textcolor{red}{Task Parsing Workflow:}

\textcolor{blue}{1. Task Parsing Agent:} "

**Genre: Science Fiction**

**Setting/Condition: A Floating City Above a Dying Earth**

**Creative Writing Prompt:** In the year 2145, humanity has retreated to a sprawling floating city known as Aetheris, suspended high above the ravaged surface of a dying Earth. The city is powered by advanced technology that harnesses the energy of storms and the sun, but resources are dwindling, and the inhabitants are beginning to feel the strain of isolation. 

 As a member of the Council of Innovators, you are tasked with solving the city's most pressing problem: how to sustain life in Aetheris while finding a way to restore the Earth below. One day, you discover an ancient artifact buried in the archives of the city\textbackslash u2014 a mysterious device that seems to pulse with energy and contains cryptic symbols. 

 Write a story exploring your character's journey as they decipher the artifact's secrets, navigate the political tensions within the council, and confront the ethical dilemmas of using the device. Will it lead to salvation for both the floating city and the Earth, or will it unleash unforeseen consequences? 

 Consider the implications of technology, the nature of survival, and the relationship between humanity and the environment as you craft your narrative."
 $\oplus$ \{ \textcolor{cyan}{TASK PROMPT}\}

\textcolor{blue}{2. Code Generation Agent:} 
"When faced with a mutation question like the one you've provided, individuals who excel in creative thinking typically approach it in several ways:

1). Understanding the Problem: They start by thoroughly understanding the existing code and its purpose. In this case, the code reads a number of test cases and computes the square of each number.

2). Identifying Opportunities for Improvement: They look for ways to enhance the functionality or efficiency of the code. For instance, they might consider:
\begin{itemize}
    \item Adding error handling for invalid inputs.
    \item Allowing for different mathematical operations (not just squaring).
    \item Implementing a more flexible input method (e.g., reading from a file or allowing for different data types).
\end{itemize}

3). Exploring Alternative Solutions: Creative thinkers often brainstorm alternative approaches to solve the same problem. They might consider:
\begin{itemize}
    \item Using a list comprehension for more concise code.
    \item Implementing a function to handle different operations based on user input.   
\end{itemize}

4). Testing and Validation: They would think about how to validate the outputs and ensure the code behaves as expected under various conditions.

5). Refactoring for Clarity: They might refactor the code to improve readability and maintainability, such as by breaking it into smaller functions or adding comments.

6). Considering Edge Cases: They would think about edge cases, such as what happens if the input is zero, negative numbers, or non-integer values.

Here's an example of how the original code could be modified to incorporate some of these creative thinking strategies:

\begin{lstlisting}
def square_number(number):
    """Returns the square of the given number."""
    return number ** 2

def main():
    import sys
    input = sys.stdin.read
    data = input().strip().splitlines()
    try:
        t = int(data[0])
    except ValueError:
        print("The first line must be an integer representing the number of test cases.")
        return
    results = []
    for i in range(1, t + 1):
        try:
            number = int(data[i])
            results.append(square_number(number))
        except ValueError:
            print(f"Invalid input at line {i + 1}: '{data[i]}'. Please enter an integer.")
            continue
    for result in results:
        print(result)

if __name__ == "__main__":
    main()
\end{lstlisting}
Key Changes Made:

\begin{itemize}
    \item Function Extraction: The squaring logic is moved to a separate function for clarity.
    \item  Error Handling: Added error handling for both the number of test cases and individual inputs.
    \item User Feedback: Provided feedback for invalid inputs to guide the user. This approach not only maintains the original functionality but also enhances the robustness and user-friendliness of the code.
\end{itemize}
"
$\oplus$ \{ \textcolor{cyan}{Output from Task Parsing Agent}\}

\textcolor{red}{Code Rewriting Workflow:}

\textcolor{blue}{1. Code Generation Agent:}
"Certainly! Here’s a mutator prompt designed to inspire creativity and innovation:

---

**Mutator Prompt: \"The Elemental Fusion\"**

Imagine a world where the four classical elements\u2014Earth, Water, Air, and Fire\u2014can be combined in unconventional ways to create new entities, phenomena, or experiences. Your task is to explore the fusion of these elements and generate innovative concepts that transcend their traditional boundaries.

**Instructions:**

1. **Elemental Combinations:** Choose two or more elements from Earth, Water, Air, and Fire. Describe how they interact and what new properties or entities emerge from their fusion. 

2. **Conceptual Expansion:** For each combination, think about how these new entities could influence the environment, society, or technology. What challenges or benefits do they bring?

3. **Unconventional Applications:** Consider how these fusions could be applied in real-world scenarios. This could include art, architecture, environmental solutions, or even new forms of entertainment.

4. **Visual Representation:** If possible, sketch or describe a visual representation of your fused element or concept. How does it look? What colors, shapes, or movements define it?

5. **Narrative Element:** Create a short story or scenario that features your new entity in action. How does it interact with the world around it? What conflicts or resolutions arise from its presence?

---

**Example:**

\begin{itemize}
    \item **Combination:** Fire + Water
    \item **New Entity:** Steam Elementals
    \item **Description:** These beings are composed of swirling steam, capable of manipulating temperature and humidity. They can create fog to obscure vision or unleash scalding vapor as a defense mechanism.
    \item **Impact:** They could help regulate climate in urban areas, but their unpredictable nature might lead to sudden weather changes.
    \item **Application:** Used in art installations to create immersive environments that change with audience interaction.
    \item **Visual Representation:** Imagine a swirling mass of steam with glowing embers, shifting in color from blue to orange.
    \item **Narrative Element:** In a city plagued by drought, the Steam Elementals emerge to restore balance, but their presence stirs fear among the inhabitants who misunderstand their intentions.
\end{itemize}

---

Feel free to explore this prompt and let your imagination run wild! What new possibilities will you uncover?"
 $\oplus$ \{ \textcolor{cyan}{TASK PROMPT}\}

\textcolor{blue}{2. Code Reviewer Agent:}"

1) Problem Description: In a world where creativity knows no bounds, we seek to develop a mutator prompt that inspires individuals to break free from conventional thinking and explore innovative mutations in their ideas, projects, or artistic expressions. The goal is to create a prompt that encourages users to think divergently, combining unrelated concepts, altering existing frameworks, and generating novel solutions or artistic expressions. This prompt should serve as a catalyst for imagination, pushing the boundaries of what is considered possible.

2) Generated Code:
\begin{lstlisting}
```python
import random

def generate_mutator_prompt():
    # Lists of unconventional concepts, actions, and mediums
    concepts = [
        \"time travel\", \"bioluminescence\", \"quantum entanglement\", 
        \"dreamscapes\", \"mythical creatures\", \"urban legends\", 
        \"chaos theory\", \"synesthesia\", \"artificial intelligence\", 
        \"ancient civilizations\", \"futuristic technology\", \"empathy\"
    ]
    
    actions = [
        \"merge\", \"transform\", \"deconstruct\", \"reimagine\", 
        \"collaborate with\", \"disrupt\", \"amplify\", \"diminish\", 
        \"animate\", \"synthesize\", \"repurpose\", \"revolutionize\"
    ]
    
    mediums = [
        \"digital art\", \"poetry\", \"interactive installations\", 
        \"performance art\", \"short stories\", \"music compositions\", 
        \"sculptures\", \"video games\", \"fashion design\", \"cooking\"
    ]
    
    # Randomly select elements from each list
    concept = random.choice(concepts)
    action = random.choice(actions)
    medium = random.choice(mediums)
    
    # Construct the mutator prompt
    prompt = f\"Imagine a world where you can {action} {concept} through {medium}. \" \\
             f\"What new realities, stories, or experiences can you create?\"
    
    return prompt

# Example usage
if __name__ == \"__main__\":
    for _ in range(5):
        print(generate_mutator_prompt())
```
\end{lstlisting}

3) Explanation of the Code: 

i. **Concepts, Actions, and Mediums**: The code defines three lists containing unconventional concepts, actions, and artistic mediums. These lists are designed to inspire creativity by combining elements that are typically not associated with each other.

ii. **Random Selection**: The `random.choice()` function is used to select one item from each list, ensuring that each prompt generated is unique and unpredictable.

iii. **Prompt Construction**: The selected elements are then combined into a coherent prompt that encourages the user to think creatively about how to merge the chosen concept with the selected action and medium.

iv. **Example Usage**: The code includes a loop that generates and prints five different mutator prompts, showcasing the variety of creative possibilities that can arise from this approach.

This mutator prompt serves as a powerful tool for artists, writers, and innovators, inviting them to explore new dimensions of creativity and push the boundaries of their imagination.

" $\oplus$ \{ \textcolor{cyan}{Output from Code Generation Agent}\}

\textcolor{blue}{3. Code Rewriting Agent:}"

**Mutator Prompt: \"The Infinite Canvas\"**

**Problem Description:**
Imagine a world where the laws of physics and logic are mere suggestions, and creativity reigns supreme. Your task is to create a Python program that generates a unique piece of art or a story based on a set of unconventional inputs. These inputs can be anything from random words, emotions, colors, or even abstract concepts. The program should utilize these inputs to produce an output that defies traditional boundaries, encouraging users to explore new dimensions of creativity.

**Comments on Previous Code:**

1. **Input Handling:** Ensure that the program can accept a diverse range of inputs, including strings, lists, and even user-generated content. This flexibility will allow for a broader spectrum of creative outputs.
   
2. **Randomization:** Incorporate elements of randomness to ensure that each execution of the program yields a different result. This could involve using random selections from lists of words, colors, or themes.

3. **Artistic Generation:** Depending on the chosen output type (art or story), utilize libraries such as `matplotlib` for visual art or `nltk` for text generation. The goal is to create something that feels alive and dynamic.

4. **User Interaction:** Consider adding an interactive component where users can tweak parameters or provide additional inputs mid-execution, allowing them to steer the creative process.

5. **Output Presentation:** Design the output to be visually appealing or narratively engaging. For art, this could mean creating a colorful canvas; for stories, it could involve formatting the text in a captivating way.

6. **Documentation:** Ensure that the code is well-commented and documented, explaining the purpose of each function and how users can modify inputs to explore different creative avenues.

---

**Example Python Code:**

Here's a simple example to get you started with the concept:

\begin{lstlisting}

```python
import random
import matplotlib.pyplot as plt
import numpy as np

def generate_random_words(num_words):
    words = [\"dream\", \"whisper\", \"ocean\", \"fire\", \"light\", \"shadow\", \"dance\", \"echo\"]
    return random.sample(words, num_words)

def create_art(words):
    plt.figure(figsize=(10, 10))
    for word in words:
        x = np.random.rand(10)
        y = np.random.rand(10)
        plt.scatter(x, y, label=word, alpha=0.5)
    plt.title(\"Art Inspired by Words\")
    plt.legend()
    plt.show()

def main():
    num_words = int(input(\"How many random words would you like to generate? \"))
    random_words = generate_random_words(num_words)
    print(\"Generated Words:\", random_words)
    create_art(random_words)

if __name__ == \"__main__\":
    main()
```
\end{lstlisting}

This code generates a set of random words and uses them to create a scatter plot, illustrating the concept of transforming abstract ideas into visual art. Feel free to expand upon this foundation, adding more complexity and creativity to the outputs!"
$\oplus$ \{ \textcolor{cyan}{Output from Code Reviewer Agent}\}
\end{tcolorbox}

\subsection{Examples of Failed Python Workflows}\label{sec:fail}

\textcolor{red}{Reason:} The ending agent of the workflow below is a code reviewer agent instead of a coding agent; hence, the final output cannot solve the coding task. 
\begin{samepage}
\begin{lstlisting}[language=Python,breaklines=true,showstringspaces=false]
steps = [
    {'name': 'task_parsing_agent', 'args': ['task_description'], 'output': 'parsed_task'},
    {'name': 'task_refinement_agent', 'args': ['task_description', 'parsed_task'], 'output': 'refined_task'},
    {'name': 'code_generation_agent', 'args': ['refined_task'], 'output': 'generated_code'},
    {'name': 'code_reviewer_agent', 'args': ['refined_task', 'generated_code'], 'output': 'review_comments'}
    ]
\end{lstlisting}
\end{samepage}

\textcolor{red}{Reason:} The input of the fortune telling agent does not match the output of the code generation agent; hence the fortune telling agent cannot fetch its input. 
\begin{samepage}
\begin{lstlisting}[language=Python,breaklines=true,showstringspaces=false]
steps = [
    {'name': 'task_parsing_agent', 'args': ['task_description'], 'output': 'parsed_task'},
    {'name': 'task_refinement_agent', 'args': ['task_description', 'parsed_task'], 'output': 'refined_task'},
    {'name': 'code_generation_agent', 'args': ['refined_task'], 'output': 'generated_code'},
    {'name': 'fortune_telling_agent', 'args':['refined_code'], 'output': 'fortune_prediction'},
    {'name': 'code_reviewer_agent', 'args': ['refined_task', 'generated_code'], 'output': 'review_comments'},
    {'name': 'code_refinement_agent', 'args': ['refined_task', 'review_comments'], 'output': 'refined_code'},
]
\end{lstlisting}
\end{samepage}

\begin{table}[h]
    \centering
    \resizebox{0.48\textwidth}{!}{
    \begin{tabular}{lccc}
        \toprule
        Total Tokens & HumanEval+ & MBPP & LCB \\
        \midrule
        Claude Opus 4.6 & 318,742 & 776,913 & 1,294,508\\
        Task Parsing Workflow & 598,361& 1,042,775& 2,031,644\\
        Code Rewriting Workflow &913,527& 1,512,908& 2,486,193\\
        \bottomrule
    \end{tabular}}
    \caption{The total tokens of the backbone model GPT-4o mini and \OURS{} generated workflows across three datasets. The total tokens include the input/output tokens on both the validation and testing sets.}
    \label{tab:cost}
\end{table}

\subsection{Token costs}\label{sec:token}
In this section, we report the token costs of the single agent (only use the backbone model), the Task Parsing Workflow and the Code Rewriting Workflow generated by \OURS{} (see Table~\ref{tab:cost}). From this table, we can learn that LiveCodeBench is a much more challenges benchmark than HumanEval+ and MBPP because the tokens cost per instance for LCB is much larger than that of MBPP and HumanEval+. In addition, the tokens cost for both workflows is larger than that for GPT-4o mini since workflows involve multiple agents taking input and output concurrently, where the Code Rewriting Workflow costs especially more tokens because it may take multiple rounds for the generator agent to produce codes passing the reviewer agent.

\end{document}